# Poisson's process in the propagation of magnetic domain wall in perpendicularly magnetized film

T. Xing,[1,2] N. Vernier,[3,2*] X. Y. Zhang,[1] Y. G. Zhang,[1] W. S. Zhao[1]

[1] *Fert Beijing Institute, MIIT Key Laboratory of Spintronics, School of Integrated Circuit Science and Engineering, Beihang University, Beijing 100191, China*

[2] *Centre de Nanosciences et de Nanotechnologies, CNRS, Université Paris-Saclay, 91120 Palaiseau, France*

[3] *Laboratoire Lumière, Matière et Interfaces, Université Paris-Saclay, 91405 Orsay, France*

We present here a statistical study of the transit time required for a magnetic domain wall to go through a small laser spot focused on 2D magnetic thin film. The domain wall velocity deduced this way is in good agreement with the other ways used to measure this parameter. But, the main fact is that the transit time is not a reproducible parameter, we have observed a quite large distribution of this parameter. This distribution can be explained assuming the movement to occur through jumps, whose probabilities are given by a Poisson's process. The fitting of this distribution has enabled us to get the required number of jumps to reverse the magnetization of the small area under the laser spot. This important parameter should lead to a better understanding of the creep regime.

Dynamic domain wall (DW) interfaces are the result of complex interactions between disorder, elasticity and thermal fluctuations [1-9]. In the low velocity limit, it has been successfully described by the creep theory, which can apply to the propagation of any interface in a medium, such as the boundaries of a crystal growing in a liquid, a fire front or the expansion of a magnetic domain [10,11]. Roughly, creep theory can be described as stochastic successive jumps of the interface in a disordered medium with many defects acting as pinning points for the propagating interface. Due to thermal activation, a pinning point can be overcome, generating a jump to reach the next pinning points [11-13]. Creep theory gives an average propagation velocity: indeed, the interface is rough and the propagation length induced by a magnetic field pulse is not the same along the interface. One has to do an averaging along the interface to check the prediction of the creep theory [7,8,12,14]. At last, most of the experiments rely on a stroboscopic procedure, using a Kerr picture before and after the pulse to view the position of the DW before and after the magnetic field pulse. Using such a way, there is no information about what is happening in real time.

The stroboscopic way is not the only way, there are few DW studies giving some clues about the real-time evolution. Several ways have been used: checking of the extraordinary Hall effect on a wire at the position of a Hall cross [15-17], pulsed lightning using a picosecond laser, which requires to do the experiment many times and assumes it works each time the same way [18] or monitoring of the Kerr signal below a laser spot focused on the sample [19-21]. But all these experiments were made using nanowires, not full film and the behavior might not be a true intrinsic 2D one.

In this paper, we report an experimental work of DW movement with real-time detection in a true 2D magnetic film of Ta(5 nm)/Co$_{40}$Fe$_{40}$B$_{20}$(1.1 nm)/MgO(1 nm)/Ta(5 nm) film with perpendicular anisotropy, already studied in a previous paper [22,23]. A linearly polarized laser spot was focused on the sample, and the reversal of the area under the spot was detected real-time through the Kerr signal S(t) [19,24]. The laser beam was arriving perpendicularly to the substrate and the configuration is a pure polar one, which probes only the perpendicular component of the film. The laser spot was approximately Gaussian, with a diameter of 2$\omega_0$ (1/e$^2$ light intensity), which could be set from 5 to 60 µm by using an additional lens between the laser and the objective lens. In the following, most of the measurements were done with 6.5 µm. The laser power was 4.5 mW to ensure negligible local heating [23]. The time response of the detector was limited by an amplifier with a bandwidth of 100 kHz, enabling us to view the rise time as 2.4 µs (time period between 10% and 90% of maximum amplitude).

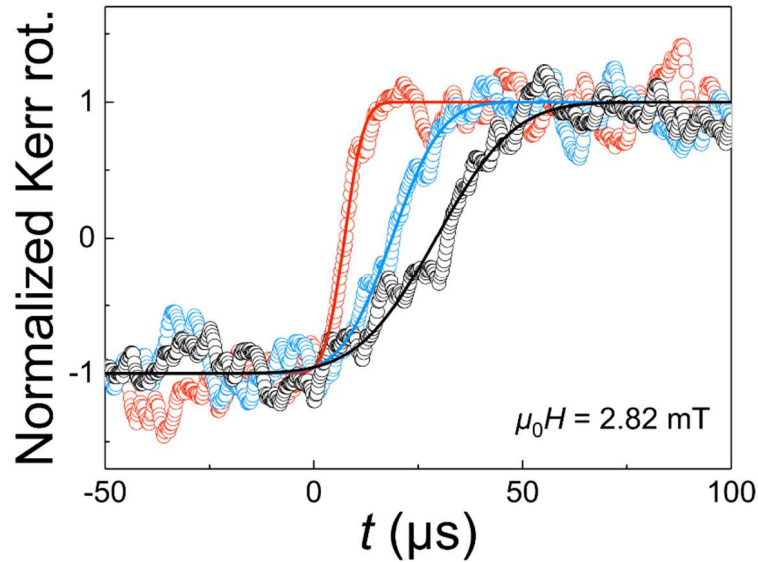

FIG. 1 : Three typical cases for different transit times with spot width 6.5 µm.

Figure 1 shows the corresponding MOKE signal when a DW goes through the laser spot driven by a 2.82 mT out of plane field. For a Gaussian spot, assuming the DW to be a straight line moving at a constant velocity, this signal is predicted to have the behavior following the error function [19,25] :

$$S(t) = A.\text{erf}[k(t-t_0)] + C \quad (1)$$

where erf(x) is the error function:

$$\text{erf}(x) = \frac{2}{\sqrt{\pi}} \int_0^x \exp(-u^2) du \quad (2)$$

The transit time $t_{tr}$ through the spot is defined as the time required for the domain to move $2\omega_0$ further. It can be deduced directly from the fitting parameter k according to:

$$t_{tr} = \frac{2\sqrt{2}}{k} \quad (3)$$

Here, it is important to point out that our experiment enables us to get the transit time of one-shot experiment. So, we have first checked if the transit time was reproducible from one attempt to another, applying exactly the same magnetic field. On a big scale, average velocity is quite well reproducible, while on a small spot of diameter 6.5 µm, it is no more true. Figure 1 presents the results, we have kept only 3 typical attempts, which gave an obviously different result, with transit times of 17.6, 33.4 and 60 µs. At this scale, we can see some stochasticity in the process of the propagation. This interesting phenomenon has already been found by some other authors [20,21].

We have used this stochastic behavior as a tool to get some information about the propagation process. First, we have determined experimentally the distribution of transit time. For one set of parameters, i.e., one value of the magnetic field, one size of the spot and one position, we have performed 100 times the experiment. A typical histogram of transit time arising from a set of experiments has been plotted in figure 2. In figure 1, the ratio signal over noise induces an error bar meaningfully lower than the step of the histogram and it allows a reliable histogram. Let's note that the size of the spot seems to be already big enough to get similar properties at any position on the sample. For the field of 1.90 mT, we have done the experiment at 4 different positions: the 4 histograms were identical within the error bars [see Supplemental Material (SM) to view these histograms] [23]. This is a quite important result, it shows that the stochastic behavior is not due to few specific defects in one area, it is an intrinsic property and, although quite small, the spot is already big enough to get a distribution of defects similar at any location on the sample. Note that, in figure 2, we have gathered altogether the results obtained at the 4 positions for 1.90 mT to get a better statistic, the four histograms can be seen in the SM [23].

Then, to fit these histograms, we have used the following model: 1) we have simplified things by assuming the meaningful defects are identical, 2) DW propagates by jumps from one pinning defect to the next one, 3) the probability of a jump to occur during a short time interval $dt$ is proportional to $dt$, this probability can be written as $dt/\tau$, $\tau$ being the average waiting time between two jumps and 4) the duration of a jump is negligible as compared to the waiting time at the pinning defects. As the density of defects is quite well defined, the number of jumps required to pass through the laser spot must be also well defined. As a result, the probability of having a transit time of duration $t$ in the spot identifies to the probability of getting n jumps during t, where n is the required number of jumps to go through the laser spot. This is a Poisson's process, and this probability is given by [26] :

$$P_n(t) = \frac{t^{n-1}}{(n-1)!\tau^n} \exp\left(-\frac{t}{\tau}\right) \quad (4)$$

Note that $P_n(t)$ is normalized. So, to fit the histograms, this function has to be multiplied by the number $N$ of sampling done. Here, usually $N = 100$. In addition, the step time $\delta T$ between two successive channels of the histograms is always small enough to have roughly a probability on one channel of the histograms equal to $P_n(t)\delta T$. As a result, the function used to fit the histograms was:

$$f(t) = NP_n(t)\delta T \qquad (5)$$

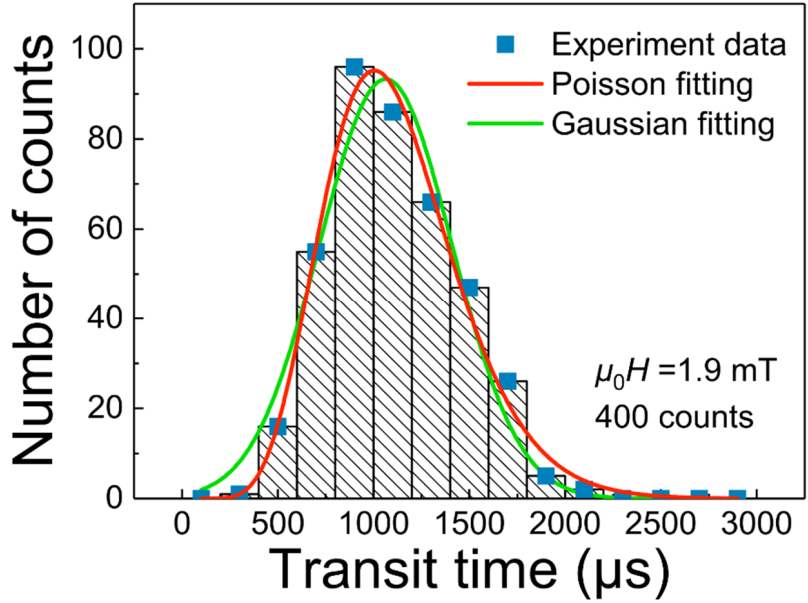

FIG. 2 : Histograms of transit time frequency distributions using the Gaussian fit (green solid line) compared to the Poisson fit (red solid line).

The result of the fit can be seen on figure 2. As a reference the fit using the usual Gaussian function has also been shown. Both fits are very satisfying, but the Poison's fit is slightly better, it explains the slight asymmetry of the histogram. Here, it must be added that, depending on the parameters (amplitude of the applied field, size of the laser spot), we have made more than 15 histograms [see SM [23]]: each time, an asymmetry could be seen and Poisson's fit appeared to be slightly better than Gaussian's ones. So, the asymmetry is meaningful and Poisson's law seems to be a good model to describe our experiment.

The small difference between Gaussian and Poisson can be explained by the $n$ values. Indeed, when $n$ is bigger than 10, Poisson's law is not very different from Gaussian's law. In fact, using a Gaussian fit instead of Poisson's one, one can get also the right parameters:

$$f(t) = \frac{N\delta T}{\sqrt{2\pi(n-1)}\ \tau} \exp\left(-\frac{(t-n\tau)^2}{2(n-1)\tau^2}\right) \qquad (6)$$

Now, we come to the most interesting result: from these fits, we can deduce how many jumps have to occur for the DW to go through the laser spot. For a laser spot of diameter 6.5 μm, $n$ has been plotted as function of the applied magnetic field in figure 3. In this range of field, within the error bars, $n$ does not change, it is around 11. So, one jump would mean a reversal of an average area of $3\mu m^2$. Let's note that one jump would imply a movement of some 1 μm, which is quite much bigger than the Larkin length (usually around 100 nm [8]). At the present time, it cannot yet be completely excluded that one main jump is in fact an avalanche process, with many smaller jumps occurring very quickly. However, elementary movements are expected to be meaningfully bigger than the Larkin length, and the jumps found out here might be elementary ones. Other references are pointing to such a size for elementary jumps [11,27,28]

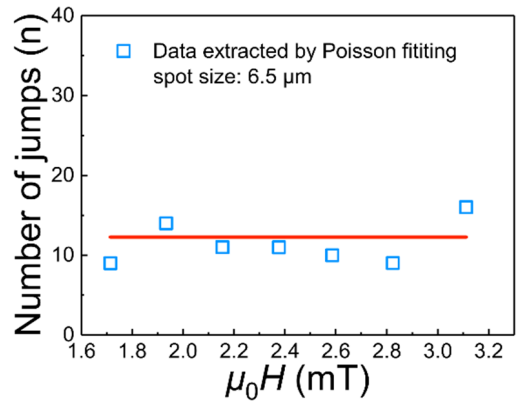

FIG. 3 : number of jumps (n) extracted by Poisson distribution fitting.

This result can be compared to propagation in nanowires: for narrow nanowires of width below one or two micrometers, some single pinning defects can stop completely the propagation of a DW [29], some discrepancies in the creep law have been found [30]. It agrees with the scale of one jump, as we can expect problems to appear when the jump area covers the full width of the wire.

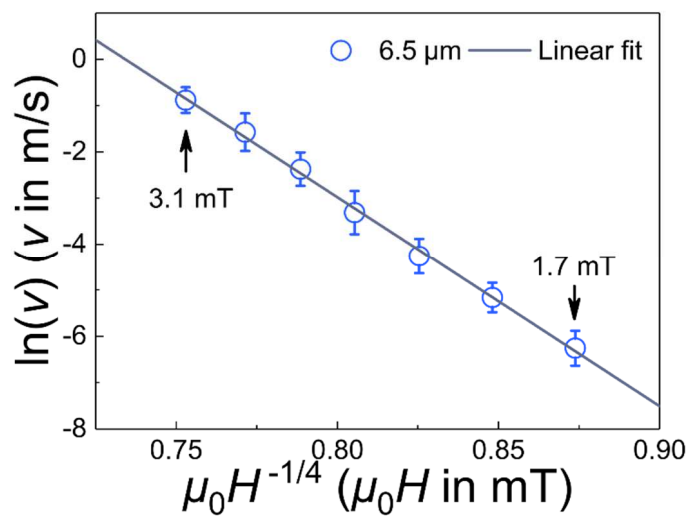

FIG. 4 : DW propagation velocity as a function of the applied magnetic field, plotted to check the creep behaviour.

The fact that $n$ does not change on the whole range of field has to be analyzed: indeed, when the magnetic field is risen up, one expects the Zeeman energy to override the potential barrier keeping the DW pinned at the position of the defect. So, we were expecting the number $n$ to decrease with the increase of the magnetic field amplitude. But, it wasn't. We think we didn't reach fields high enough to reach the possibility of overriding any main defect potential. So, whatever the field was, pinning was occurring for each relevant defect and the number of jumps was the same. Indeed, when looking at propagation as a function of magnetic field (see figure 4), we can see that even at the highest applied field, we are still in the creep regime, which means pinning by defects is still working.

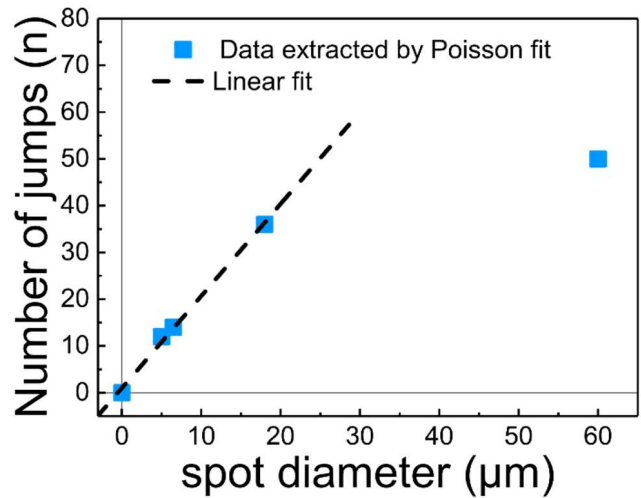

FIG. 5 : The number of jumps as a function of probing area at 1.9 mT. Note that a point at zero has been added as there is no more pinning point when the area goes to zero, which means no jump over a defects to go through.

To check our model, we have changed the size of the spot. Within the error bars, the first result is the fact the characteristic mean time $\tau$ between two jumps does not really change with the size of the spot for a given field, although the diameter of the spot goes from 5 μm to 60 μm. It seems consistent with our model, as $\tau$ is defined at a much lower scale. The second parameter which can be checked in this experiment is the number of jumps: it has been plotted in figure 5. As the area of the spot increases, as expected, the number of jumps increases. But, the dependency is not so easy to explain. Indeed, in a Markovian process, one could have expected it to increase linearly with the area of the spot, which means the square of the diameter. Obviously, it is more complicated (see DW pinning and depinning process observed with Kerr microscopy in the SM) [23]. In fact, the first idea implying the area might be wrong. Indeed, the length of the DW inside the spot might be a relevant parameter for the probability of getting one jump, as the number of possible jumps is increasing with this parameter. So, the bigger the spot is, the bigger should become the probability. Such an analysis requires first some more data as well as some more thinking and it goes beyond the aim of this paper. But, according to this idea, the number of possible jumps should increase linearly with the diameter of the spot, and we have indeed a linear behavior up to a diameter of 20 µm (figure 5).

In conclusion, we have carried a real-time detection of a magnetic DW transit in a small area defined by a laser spot. Using the same magnetic field, the transit time was not reproducible. To understand this stochasticity, we have studied the distribution of transit time obtained on a statistic of 100 experiments. To fit the histograms, we have assumed a propagation through stochastic jumps, ruled by a Poisson's statistic. We have got a very good agreement between experiments and model. In particular, the model has been able to explain the slight but systematic asymmetry of the histograms. As a result, we could get an important parameter, which is the required number of jumps of the DW to go through the spot: for a spot of diameter 6.5 µm, the fitting shows that there are around 11 jumps. It means an average reversal of a 3 µm$^2$ area during each jump. This is an important new information about the defects ruling propagation in magnetic thin films and inducing the creep behavior. Let's note that this kind of analysis can apply to any creep experiment, it is a powerful tool to get a better understanding of this phenomenon.

The authors wish to thank the support from the program of China Scholarships Council (No. 201906020020).

Supplemental material

Note 1. Sample characteristic

The Ta(5 nm)/CuN(40 nm)/Ta(5 nm)/Co40Fe40B20(1.1 nm)/MgO (1 nm)/Ta(5 nm) thin film was deposited on Si/SiO2 substrate using magnetron sputtering deposition technique (Singulus, Germany). It was annealed at 380°C for 20 minutes. Fig. S1 shows the hysteresis loop measured by polar MOKE, indicating the presence of PMA in our sample. The magnetization of the magnetic layer was detected to be $\mu_0 M_S = 1.2 \pm 0.1$ T using the method described in ref. [31].

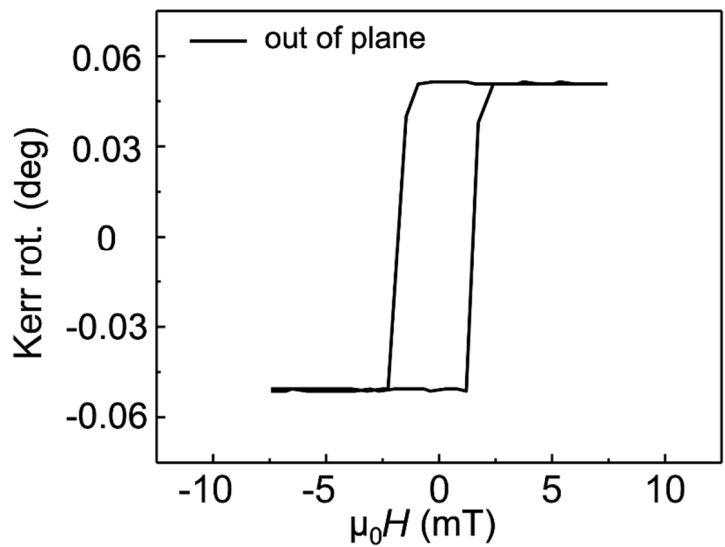

FIG. S1 Hysteresis loop measured by polar MOKE.

Note 2. local heating due to the laser spot

To evaluate the local heating due to the laser spot, we performed calculations with the following hypothesis:

- Because there is a half reflective plate, the power arriving on the sample is only P = 2.2 mW. As the sample is acting as a mirror (indeed, fortunately, there is a reflected beam, whose polarization is analyzed to determine the magnetic state due to Kerr effect), only a small part of the light is absorbed by the sample. Here, we have assumed that 20% of the power was absorbed, which is an upper limit. As shown in Fig. S2, this power is uniformly absorbed in a half sphere of radius equal to the spot size (r = R). This hypothesis can be justified by the skin effect, which means light acts only on the surface area

- The temperature is approximately uniform in the half sphere enlighten by the laser spot. It will be written as $T_1$. On the substrate far away from the laser spot, the temperature of the substrate is $T_0$.

- The energy given to the sample by the laser is removed by heat conduction through substrate.

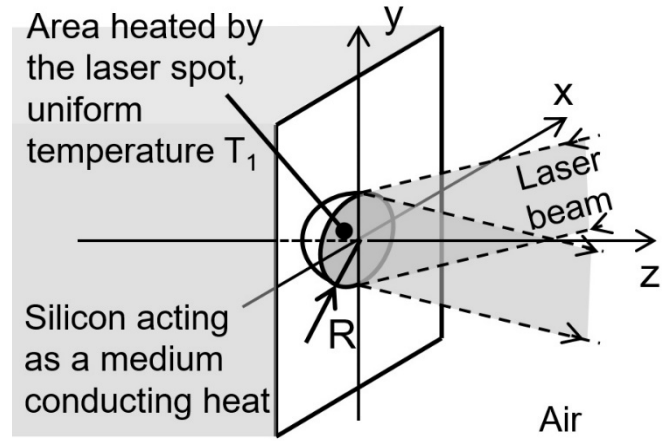

FIG. S2 Schematic of local heating due to the laser spot.

Outside, the half sphere, there is no heat source and the heat equation to solve is :

$$\Delta T = 0$$

Using spherical coordinates, the solution is:

$$T(r) = \frac{K_1}{r} + K_2$$

In steady state, boundary values are :

- r = R : heat flux = power deposited on the substrate $\Rightarrow$ $-\lambda \frac{\partial T}{\partial r} 2\pi R^2 = \varepsilon P$ $\Rightarrow$ $K_1 = \frac{\varepsilon P}{2\pi \lambda}$

- r = +∞ : T = $T_0$ $\Rightarrow$ $K_2 = T_0$

As $T_1 = T(R)$, it leads to the solution :

$$T_1 - T_0 = \frac{\varepsilon P}{2\lambda \pi R}$$

Using, $\varepsilon = 0.2$, P = 2.2 mW, $\lambda$ = 148 W.m$^{-1}$.K$^{-1}$ and R = 6.5 µm, we get $T_1 - T_0$ = 0.07 K

Let's note that's we have used an upper limit for $\varepsilon$, which means we can be sure that the heating remains below 0.1 K.

Note 3. Transit time distributions at 4 positions

We selected 4 random positions on the sample surface, which were performed by keeping the optical path fixed and moving the sample holder. Based on the defects in the yellow dashed line in Fig. S3 (a), we can align the 4 selected measurement areas and obtain their relative positions (blue solid line). The light intensity distributions at the 4 positions are essentially the same. As shown in Fig. S3 (b), we take the profile of 4 areas along the transverse direction. The fitted curves show that the four spots have the same Gaussian distribution with a spot diameter of 6.5 µm.

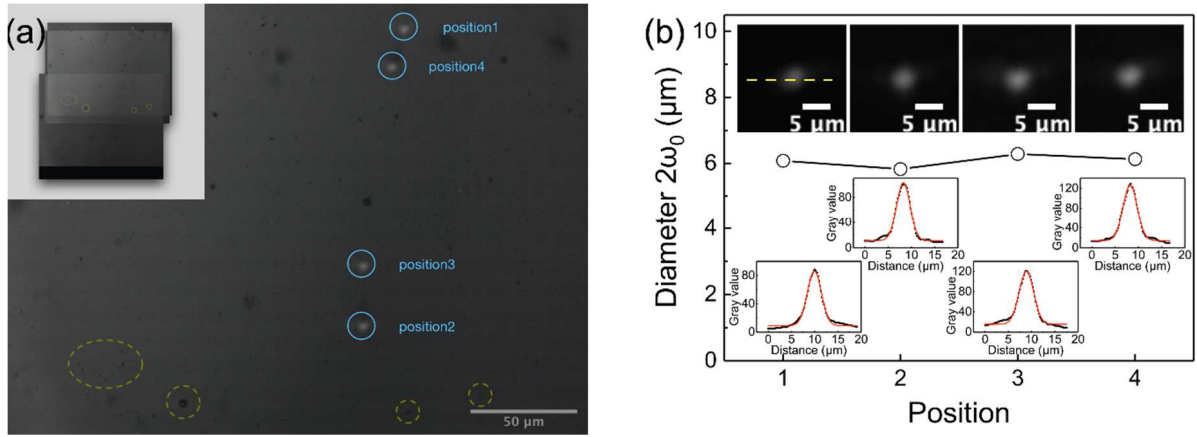

FIG. S3 (a) The relative positions of 4 spot areas. (b) Spot diameters at 4 positions and the respective Gaussian fits.

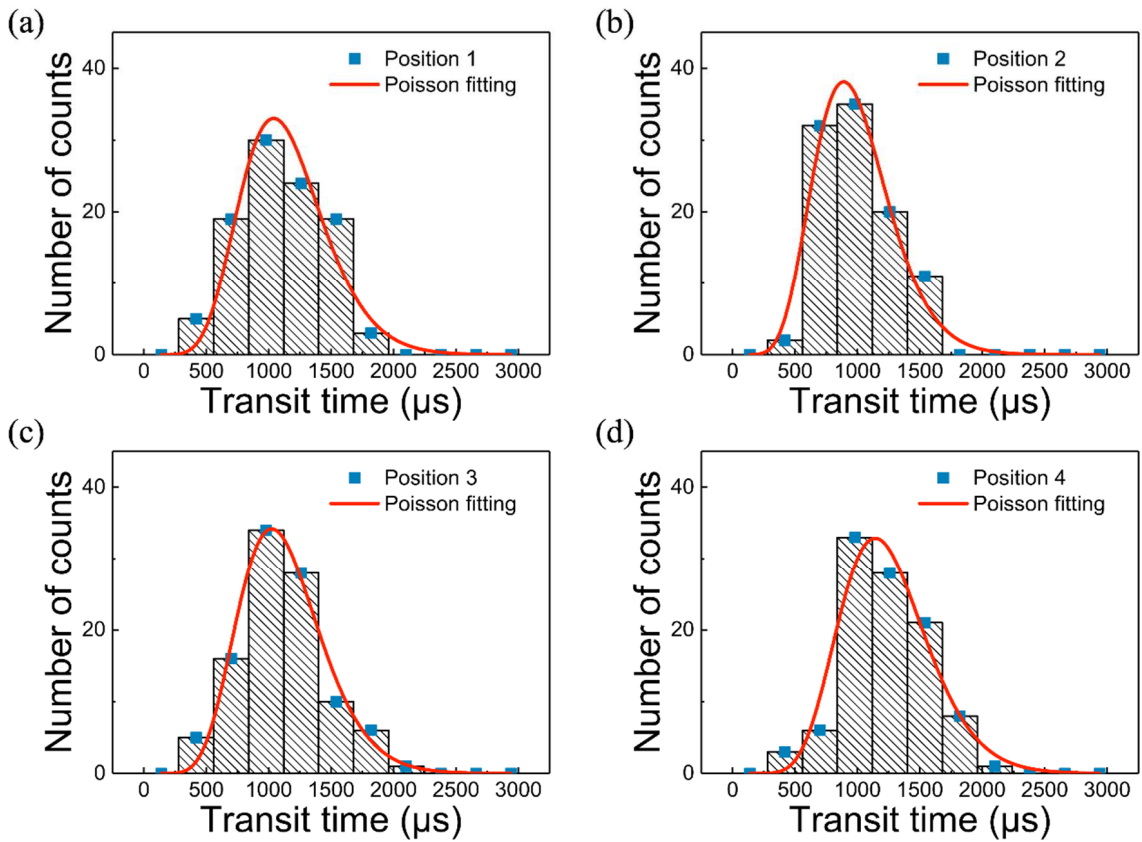

FIG. S4 Transit time distributions on (a) position 1, (b) position 2, (c) position 3 and (d) position 4.

The transit time distributions were then measured for each location, where each distribution consists of 100 counts. The histogram of counts is plotted in Fig. S4 with the histogram step of 280 µs. At each position, the histograms of counts show similar characteristics. First, the maximum values are located at about 1000µs. Then the distribution exhibits a slight central asymmetry. On both sides of the maximum value, the rate of count decrease on the left is faster than on the right. Thus, all four distributions show a shift to the left.

We also used the Poisson model to fit the transit time distributions for the four positions and plot the red solid curves in Fig. S4. To verify the intrinsic homogeneity, we compare the fitting parameters for the four positions. Table S1 lists the parameters corresponding to the positions, where histogram step is set as a free variable for comparison with known expected values. The histogram steps extracted from the fits show small relative errors from the expected values, indicating that the distribution is well described in each fit. The number of jumps and average waiting time determine the width and the asymmetry of Poisson distribution. At different positions, these parameters show very small fluctuations, proving that the 4 positions are essentially homogeneous.

In summary, we randomly selected 4 positions for measurement. The distributions show very similar properties. The homogeneity of 4 positions suggests that the distributions do not come from individual local defects, which we believe is an intrinsic property of the film.

**Table S1** Fitting parameters

| Positions | Number of jumps n | Average waiting time $\tau$ (μs) |
|---|---|---|
| Position 1 | 11 | 104 |
| Position 2 | 10 | 99 |
| Position 3 | 11 | 102 |
| Position 4 | 12 | 103 |

Note 4. Transit time distributions under different magnetic fields and spot sizes

1) Spot of diameter 6.5 µm :

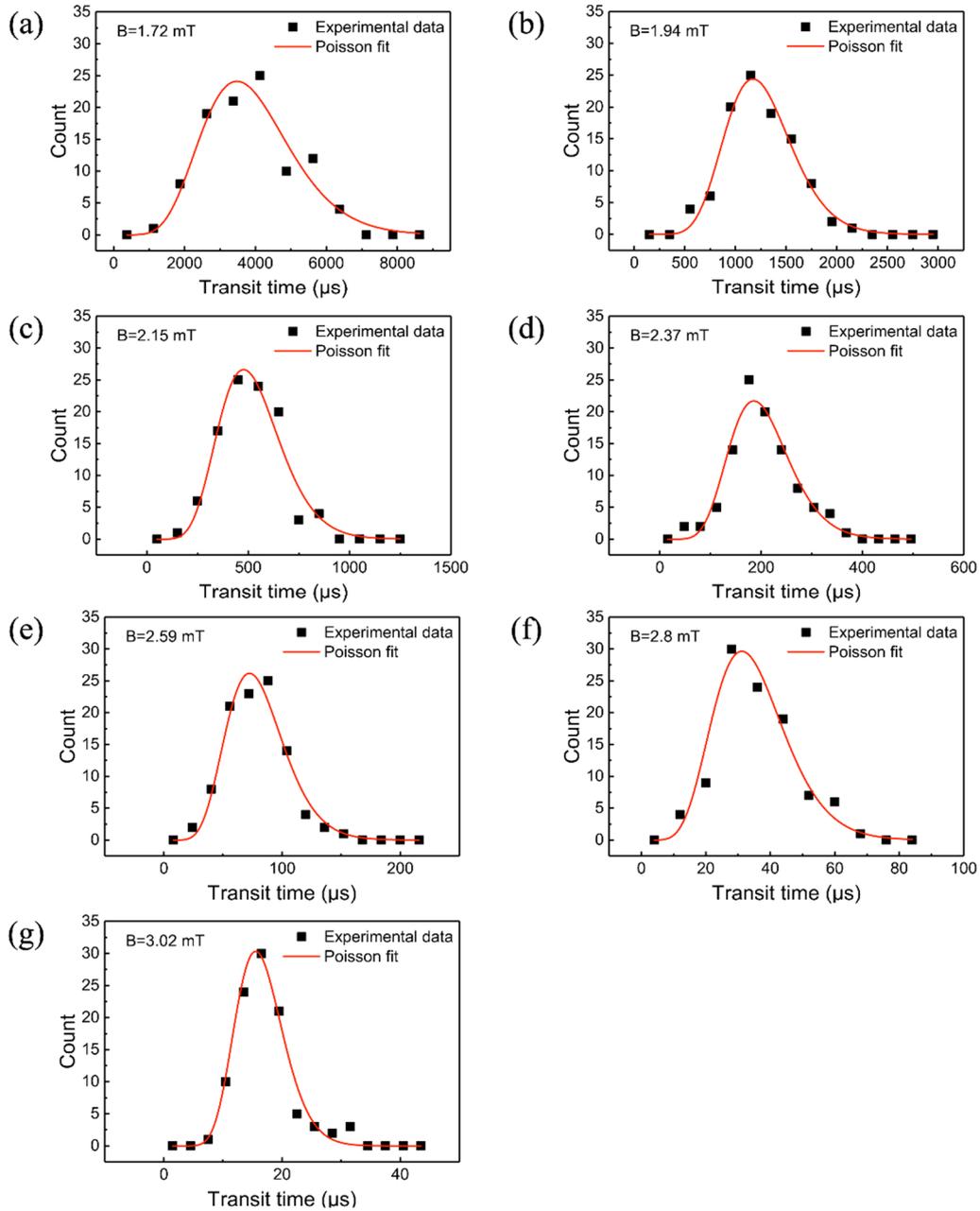

FIG. S5 (a)-(g), transit time distributions at magnetic fields from 1.72 mT to 3.02 mT using the spot of diameter 6.5 µm.

We have also measured the transit time under different magnetic fields with the spot diameter of 6.5 µm. At each magnetic field, the frequency counts of 100 statistics show the distribution property. According to the Poisson model, the red fitted curves in Figure S5 describe the frequency counts very well. It also clearly demonstrates the slight asymmetry of each distribution.

2) Spot of diameter 60 µm :

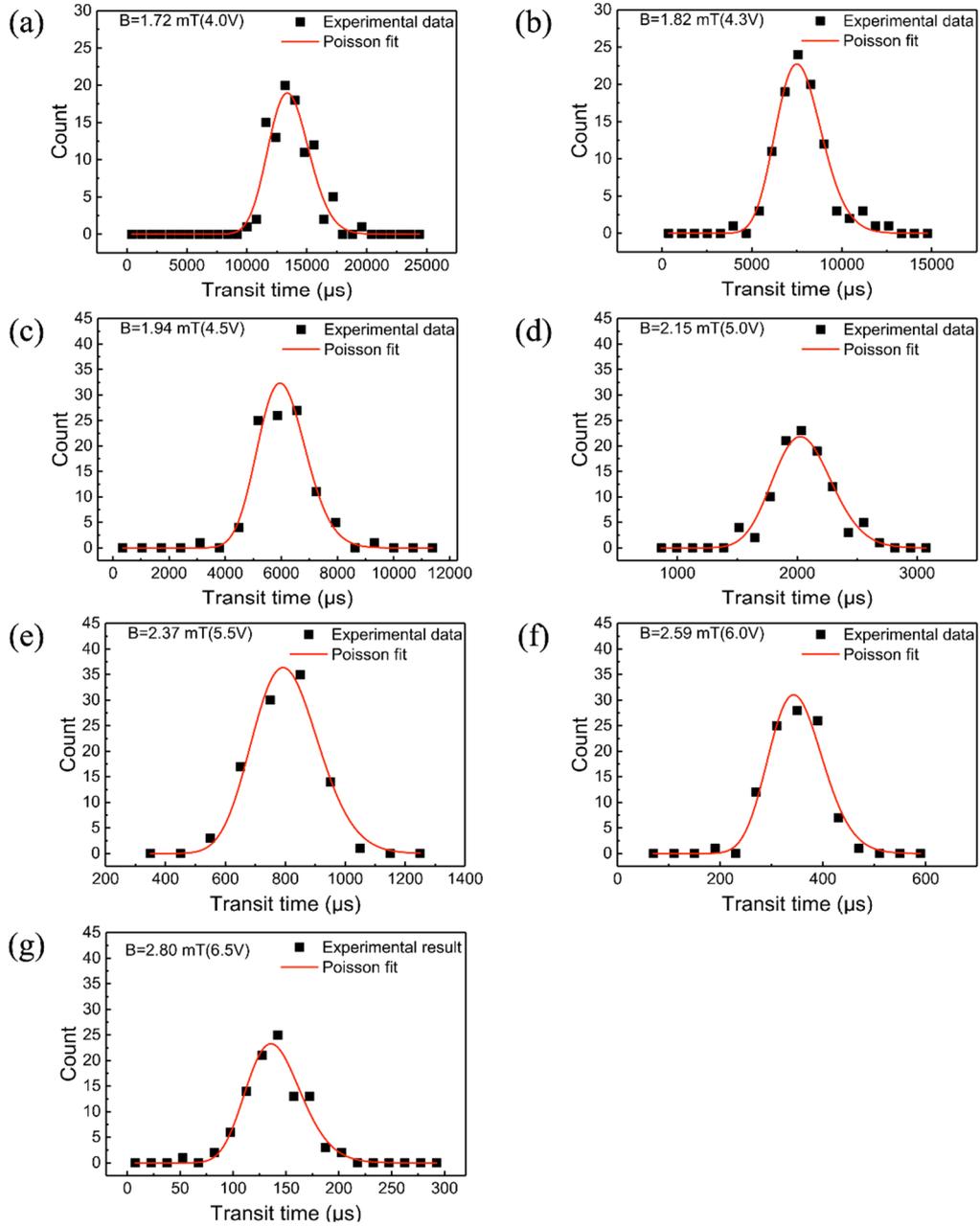

FIG. S6 (a)-(g), transit time distributions at magnetic fields from 1.72 mT to 2.8 mT using the spot of diameter 60 µm.

To check the distribution at big scale, we also performed experiments to measure the transit time at different magnetic fields with a spot diameter of 60 µm (figure S6).

3) Spot diameters of 5 and 18 µm at 1.7 mT magnetic field :

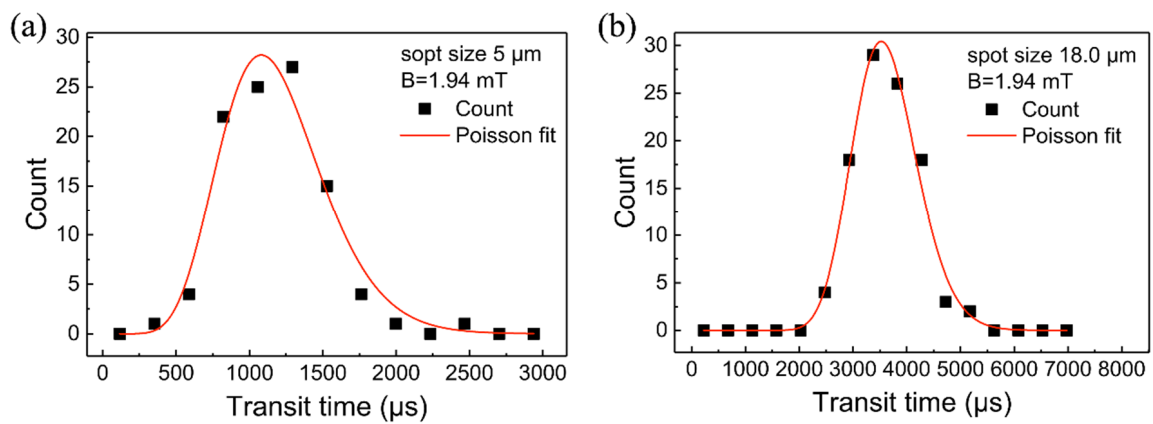

FIG. S7 Transit time distributions at magnetic fields of 1.94 mT using the spot of diameter 5 µm (a) and 18 µm (b).

To investigate more spot sizes, we performed experiments to measure the transit time at 1.94 mT magnetic field with spot diameters of 5µm and 18 µm (Figure S7).

Note 5. Domain wall pinning and depinning process observed with Kerr microscopy

The domain wall pinning and depinning process in the film is observed using a high-resolution Kerr microscopy. As marked by red circles in FIG. S8, some pining sites exist in the magnetic film and the domain wall is deformed by these pinning sites when propagating. After the domain wall is suddenly depinned from a pinning site, the magnetization in the nearby region switches rapidly. A jump of Kerr signals will take place when this process is detected by a laser beam.

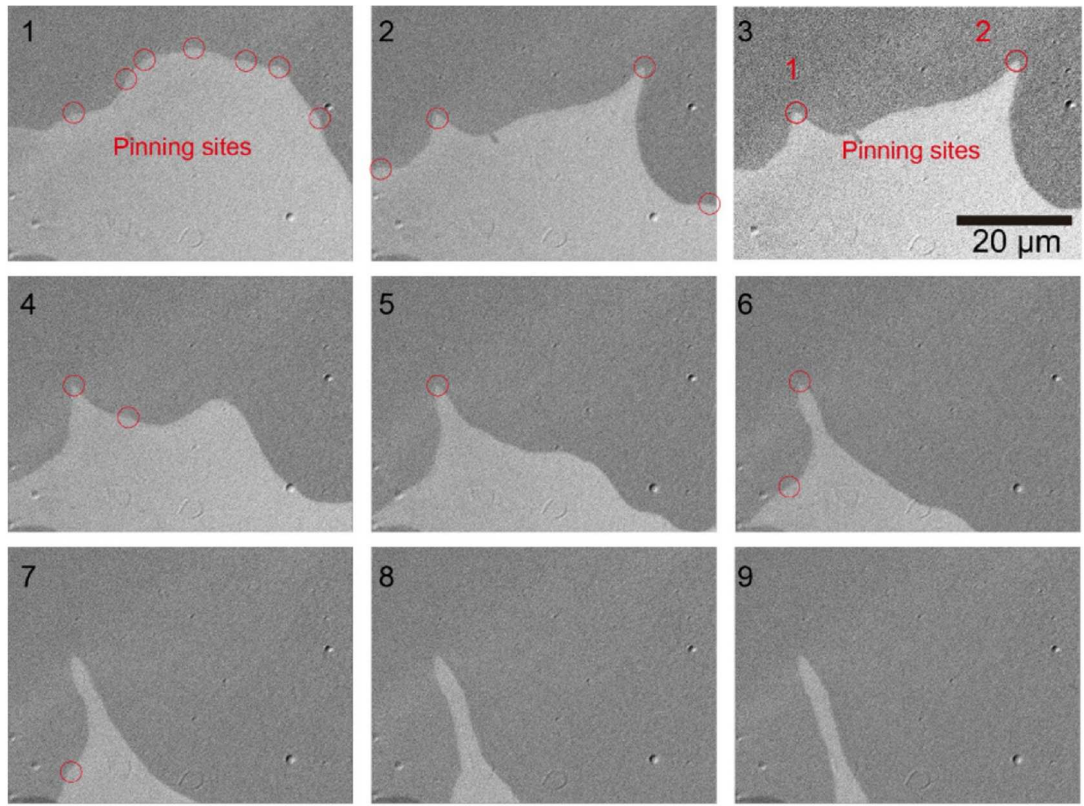

FIG. S8 Domain wall pinning and depinning process observed by using Kerr microscopy. The brightness contrast in different domains represents opposite magnetization and red circles indicate the position of pinning sites. A sequence of magnetic field pulses with 0.35 mT magnitude and 0.5s duration was applied and a Kerr image is taken after each pulse. Information of the sample: Ta(5nm)/CoFeB(1.0)/MgO(2.0nm)/Ta(5nm), annealed at 300°C for 2 hours. Note that the density a very strong defects such as defects 1 or 2 of this figure is very small : one or two in an area of 2500µm2. The odds that the spot laser (diameter 6 µm) was on such a pinning defect for the transit time measurements is very small. We guess we would have seen some very different behaviour when looking at the 4 different spots (see figure S2) if we had had such a bad luck.